\documentclass[aps,prl,twocolumn,groupedaddress,showpacs,floatfix]{revtex4}

\usepackage{graphicx}
\usepackage{dcolumn}

\begin{document}



\title{Accurate interaction energies from perturbation theory
       based on Kohn-Sham model}

\author{Rafa{\l} Podeszwa}

\affiliation{%
Department of Physics and Astronomy,
        University of Delaware,  Newark, DE 19716}

\author{Krzysztof Szalewicz}%
\affiliation{%
Department of Physics and Astronomy,
        University of Delaware,  Newark, DE 19716}
%




\date{\today}

\begin{abstract}
The density-functional based symmetry-adapted perturbation theory
[SAPT(DFT)] has been applied to the argon,  krypton,  and benzene
dimers.  It is shown that---at a small fraction of computational
costs---SAPT(DFT) can provide similar accuracies
for the interaction energies as high-level wave-function based methods
with extrapolations to the complete basis set limits.
This accuracy is significantly higher than that of any other DFT or DFT-based
approaches proposed to date. 
\end{abstract}


\maketitle



The intermolecular forces---sometimes called van der Waals (vdW) interactions
or forces---determine the structure and properties of most clusters,  liquids,
and solids.  These forces also govern many life processes,
such as the genetic code replication,  protein structure and dynamics,
and enzymatic actions.  Thus,  the ability to computationally predict
van der Waals forces is significant for the understanding of all these
systems.  However,  although the standard wave-function based electronic
structure methods can in principle be used for such predictions,
in practice these methods are too time consuming to be applied to most
systems of interest in biology,  even with extensive use of the current
computers capabilities.  The density-functional theory (DFT) methods
are much less time-consuming, 
however,  the currently existing versions of DFT fail to
describe the dispersion interaction, an important part
of the van der Waals force.  This problem is due to the fact that the
dispersion forces result from long-range correlations
between electrons,  whereas the current exchange-correlation
potentials model only short-range correlation effects.
Many authors add the asymptotic expansion of the dispersion energy
to the DFT interaction energies,  which inherently includes some double
counting (see Ref.~\onlinecite{Wu:01} for a discussion of these issues).

Occasionally,  for a specific system, one of the variants of DFT can give
reasonably good predictions of interaction energies,  at least
in some regions of a potential energy surface.  This fact encouraged 
some authors to build system-specific potentials fitted to a number
of grid points on the potential energy surface computed using a wave-function
based method.  For example,  Boese {\em et al.\/}~\cite{Boese:03} optimized
an ammonia-specific potential by adjusting some parameters in one
of the standard functionals.  Recently,  Lilienfeld {\em et al.\/}~%
\cite{Lilienfeld:04} proposed to use atom-centered nonlocal effective core
potentials with parameters adjusted for specific systems.  These methods
do not offer physically motivated improvement of the density-functional
formalism but rather rely on cancellations of errors.  Also,
a number of wave-function calculations are needed to optimize the
parameters in the functionals,  which limits the
range of applications to systems that can be treated with the
latter methods (unless the parameters can be shown to be transferable).  

An approach which uses the specific characteristic of the dispersion
interaction has recently been presented by Dion {\em et al.\/}~\cite{Dion:04}.
The method was denoted by the authors as vdW-DF.
It adds a nonlocal correlation energy part to existing functionals.
This term models the dispersion energy utilizing approximate density response
functions.  The method predicts interaction energies of the systems
investigated in Ref.~\onlinecite{Dion:04} qualitatively
(to within a factor of about 1.5, see below).

Another approach to the calculations of interaction energies for large
molecules was developed 
by Misquitta {\em et al.}~\cite{Misquitta:02,Misquitta:03} and independently
by Hesselmann and Jansen~\cite{Hesselmann:02-03},
following ideas of Williams and Chabalowski~\cite{Williams:01}.
This approach is based on symmetry-adapted perturbation theory (SAPT)
\cite{Jeziorski:94},
but utilizes the description of the interacting monomers in
terms of Kohn-Sham (KS) orbitals,  orbital energies,  and frequency-dependent
density susceptibility (FDDS) functions.  
The DFT-based SAPT will be called SAPT(DFT).  This method can be shown
to be potentially exact for all major components of the interaction energy
(asymptotically for exchange interactions)
in the sense that these components would be exact if the DFT description
of the monomers were exact~\cite{Misquitta:02,Misquitta:03,Misquitta:04}.
Applications to a number of small dimers have shown that SAPT(DFT)
provides surprisingly accurate results,  sometimes more accurate than
the standard SAPT at the currently programmed level~%
\cite{Misquitta:04,Misquitta:05}.

The regular SAPT method involves expansions in powers of the intermonomer
interaction operator $V$ and the intramonomer correlation operator $W$, the
so-called M\o ller-Plesset (MP) potential.  The terms proportional to powers
of $W$ describe the effects of intramonomer electron correlation on the
interaction energy.  Similarly as in the electronic-structure many-body
perturbation theory (MBPT) or coupled-cluster (CC) methods,
these terms are expensive to compute, with CPU times scaling as a high power
of system size $N$---the seventh power if the complete currently programmed
set of SAPT components is included.  This scaling is the same as for MBPT
in the fourth order (MP4) or CC including single,  double,  and noniterated
triple excitations [CCSD(T)].  In SAPT(DFT), no terms of this type appear,
as the intramonomer correlation effects are accounted for by DFT.
Therefore,  SAPT(DFT) scales as only $N^5$,  i.e.,
a SAPT(DFT) calculation is generally orders of magnitude
faster than a regular SAPT calculation at the complete currently
programmed level.
This computational advantage is further significantly increased by
the superior basis set convergence of SAPT(DFT) compared to 
the wave-function based electron correlation methods.  The latter
methods converge slowly due to the necessity of reproducing the intramonomer
electron-electron cusps by expansions in products of one-electron functions.
Such expansions do not appear in DFT.  In SAPT(DFT), the orbital-product
expansions are present in the expressions for the dispersion energy,  however,
it has been shown~\cite{Misquitta:04,Misquitta:05} that this component
(similarly as the SAPT dispersion energy of zeroth order in $W$)
can be saturated in reasonably small basis sets provided that ``midbond"
functions are used,  i.e.,  basis functions are placed at a point between
the two monomers.  It appears that,  in many cases,  polarized triple-zeta (TZ)
quality bases give SAPT(DFT) interaction energy components
converged to a similar number of digits as the regular SAPT
components in polarized quadruple-zeta (QZ) bases.  This results in a difference
in the basis set size of about a factor of two,  a ratio of 2$^4$ in
computer time at the MP4 or CCSD(T) level.  Our current implementation
of SAPT(DFT) is using an interface to the time-dependent DFT (TD-DFT)
part CADPAC~\cite{CADPACshort} to compute FDDS's, which is the time-limiting
step of the
calculations.  An optimized TD-DFT program now under development should
decrease time requirements of this step by at least one order of magnitude.
This will make the largest calculations described here comparable
to the supermolecular DFT calculations in the
same basis.  Thus, the SAPT(DFT) method is not prohibitively
expensive,  as stated by the authors of Ref.~\onlinecite{Lilienfeld:04},
and it has already been applied
to systems containing about 40 atoms and 200 electrons~\cite{Podeszwa:04}.

We present here the first application of SAPT(DFT) to relatively
large interacting monomers.  In order to compare with the vdW-DF method,
we have chosen the same systems as investigated in Ref.~\onlinecite{Dion:04}:
Ar$_2$, Kr$_2$,  and the benzene dimer.
We have used the following Cartesian basis sets: aug-cc-pV5Z~\cite{Woon:93} for Ar$_2$ (with
$g$ and $h$ functions removed due to restrictions of CADPAC), 
aug-cc-pVTZ~\cite{Woon:93} for Kr$_2$, and a 
polarized double-zeta size basis with polarization
coefficients optimized on dispersion energy~\cite{Bukowski:99} for benzene.
In all cases, we used a set of midbond functions 
consisting of $s$ and $p$ functions with exponents
$0.9$, $0.3$, $0.1$, and $d$ and $f$ functions with exponents
$0.6$ and $0.2$, placed at the center of mass of each dimer.
We used the monomer-centered ``plus'' form of basis sets~\cite{Williams:95}
for the argon (including $d$ functions on the interacting partner) 
and benzene dimers,
and the full dimer-centered form
Kr$_2$. For the benzene dimer, 
we considered the parallel ``sandwich'' configuration, and
the monomer geometry was taken from Ref.~\onlinecite{Tsuzuki:02}.
We employed PBE0~\cite{Adamo:99} DFT functional with the asymptotic
correction \cite{Misquitta:02} for all the systems. 
In addition, the B97-2~\cite{Wilson:01} functional was applied at near minimum 
geometries for all dimers and at all points for the argon dimer.
Effects of the third and higher orders in $V$ have been neglected.
In TD-DFT, we have used the standard PBE0 or B97-2 kernels 
for argon and krypton, and the
LDA kernel for benzene.  The use of the LDA kernel provides a considerable
speedup of calculations.  To check the accuracy of this approximation,
we performed a single-point calculation using the PBE0 kernel
for the benzene dimer.  The error in the dispersion energy was
smaller than 1\%.   

 \begin{figure}
 \includegraphics[scale=0.7]{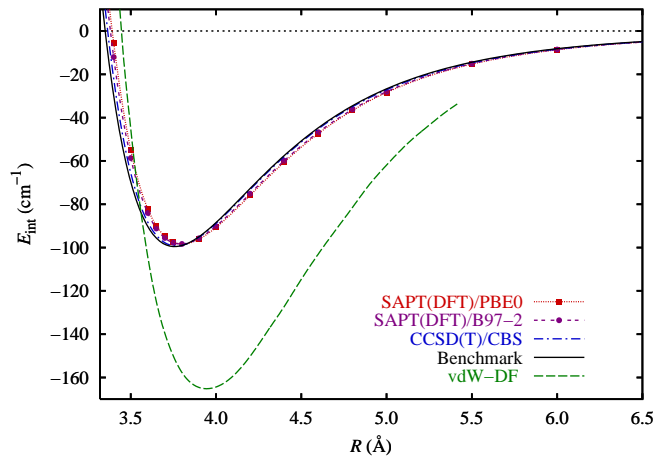}%
 \caption{\label{fig:argon} (Color online). Interaction energy of the argon dimer.
 SAPT(DFT)---this work. 
 CCSD(T)/CBS---Ref.~\onlinecite{Patkowski:04}.  
 Benchmark---Aziz, Ref.~\onlinecite{Aziz:93}.
 vdW-DF---Ref.~\onlinecite{Dion:04}.}
 \end{figure}

Figure \ref{fig:argon} presents the interaction potential for the argon
dimer.  The benchmark results are the empirical potential of Aziz
\cite{Aziz:93} and the CCSD(T) potential with extrapolation to the complete
basis set (CBS) limit by Patkowski {\em et al.}~\cite{Patkowski:04}.
The two curves are almost indistinguishable,  showing the very
high level of agreement between {\em ab initio} theory and experiment
for this system.  The SAPT(DFT) calculations are very close to the
benchmarks,  within about 2 cm$^{-1}$ or 2\% at the minimum geometry. 
In contrast,  the vdW-DF method
gives a curve which is about 1.6 times too deep,  and the minimum
position is shifted by 0.2 \AA.

 \begin{figure}
 \includegraphics[scale=0.7]{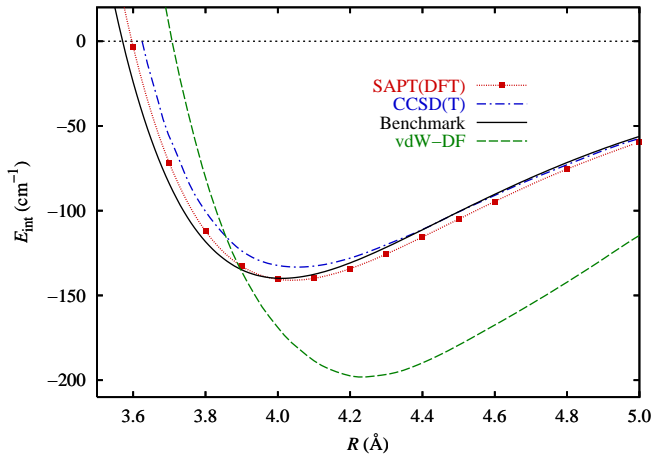}%
 \caption{\label{fig:krypton} (Color online). Interaction energy of the krypton dimer.
 SAPT(DFT)---this work. Benchmark---Dham {\em et al.\/}~\cite{Dham:89}.
 CCSD(T)---calculations in aug-cc-pV5Z+{\it spdfg} basis
 by Slav\'{\i}\v{c}ek {\em et al.\/}~\cite{Slavicek:03}.
 vdW-DF---Ref.~\onlinecite{Dion:04}.}
 \end{figure}

Our results for the krypton dimer are displayed in Fig.~\ref{fig:krypton}.
For this system, the benchmark curve is given by the empirical fit
of Dham {\em et al.\/}~\cite{Dham:89}.  The CCSD(T) curve computed by 
Slav\'{\i}\v{c}ek
{\em et al.\/}~\cite{Slavicek:03} represents the best literature 
non-relativistic theoretical
result and is in fact at the limit of what the {\em ab initio\/} methods are
capable of achieving at the present time.  The SAPT(DFT) curve agrees
with the benchmark slightly better than
the CCSD(T) one,  and much better than the curve produced by the
vdW-DF method~\cite{Dion:04}.  The latter curve is about a factor of 1.4
too deep and has the minimum position shifted by about 0.2 \AA.

 \begin{figure}
 \includegraphics[scale=0.7]{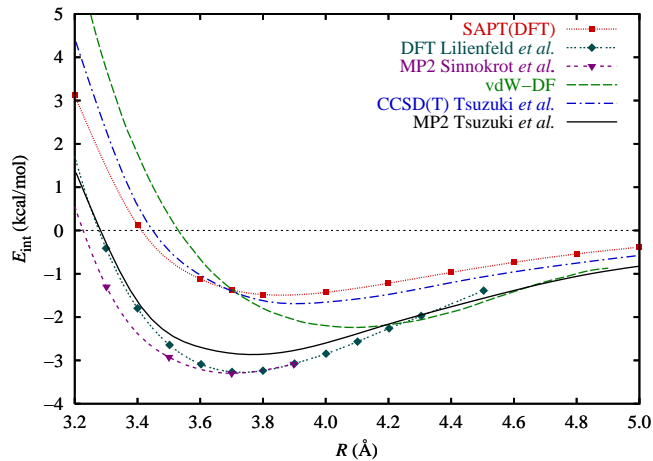}%
 \caption{\label{fig:benzene} (Color online). 
 Interaction energy between two benzene molecules
 in the parallel ``sandwich'' configuration. SAPT(DFT)---this work.
 CCSD(T)---Ref.~\onlinecite{Tsuzuki:02} (Model II).
 MP2 Tsuzuki {\em et al.\/}~\cite{Tsuzuki:02}---computed in the
 aug(d,p)-6-311G** basis.
 MP2 Sinnokrot {\em et al.\/}~\cite{Sinnokrot:02}---computed
 in the aug-cc-pVTZ basis.
 vdW-DF---Ref.~\onlinecite{Dion:04}.
 DFT Lilienfeld {\em et al.\/}---Ref.~\onlinecite{Lilienfeld:04}.
 }
 \end{figure}

 \begin{table}
 \caption{\label{tab:minima} Dimer interaction energies near minima. SAPT(DFT)
 results obtained using the PBE0 and B97-2 functionals are compared
 with benchmark results. For the argon and krypton dimers the
 energies are in $\rm cm^{-1}$, for the benzene dimer in kcal/mol. }
 \begin{ruledtabular}
 \begin{tabular}{ldddr}
   & \multicolumn{1}{c}{$R (\mbox{\AA})$} & \multicolumn{1}{c}{PBE0} & \multicolumn{1}{c}{B97-2} & \multicolumn{1}{c}{Benchmark}\\ \hline
Ar$_2$          & 3.75 & -97.46  &  -97.81   & $-99.64$\footnote{Ref.~\onlinecite{Aziz:93}.} \\
Kr$_2$          & 4.00 & -140.49 &  -141.36   & $-139.87$\footnote{Ref.~\onlinecite{Dham:89}.} \\
$\rm (C_6H_6)_2$& 3.80 &  -1.48  &  -1.52   &   \mbox{$-1.48$\footnote{Ref.~\onlinecite{Tsuzuki:02}---CCSD(T) (Model III).},
                                    $-1.81$\footnote{Ref.~\onlinecite{Sinnokrot:02}---CCSD(T).}}\\
 \end{tabular}
 \end{ruledtabular}
 \end{table}
Finally,  the benzene dimer results are presented in Fig.~\ref{fig:benzene}.
For this system,  there is no highly reliable benchmark available.
The best {\em ab initio\/} calculations for a number of 
monomer separations are those by Tsuzuki {\em et al.\/}~\cite{Tsuzuki:02} 
at the CCSD(T) level.  The values displayed
were termed as ``Model II''
and included the MP2 energies computed in a polarized DZ-quality basis
augmented with a single diffuse polarization set [
aug(d,p)-6-311G**] and the $\Delta$CCSD(T) = CCSD(T)-MP2
component computed in a DZ basis polarized only on carbons.
Below we will quote the MP2 and $\Delta$CCSD(T) values in parentheses.
Model II gives $-1.62$ ($-2.85$, $1.24$) kcal/mol at $3.8$~\AA.
The authors computed also an extrapolated interaction energy (``Model III",
only at $3.8$~\AA) by performing a CBS extrapolation at the MP2 level and
scaling the $\Delta$CCSD(T) component,  which gave $-1.48$ ($-3.28$, $1.80$)
kcal/mol.
%
Accurate calculations for the benzene dimer were also performed
by Sinnokrot {\em et al.\/}~\cite{Sinnokrot:02},  but only at a single point,
$R$ = $3.7$ \AA.  These authors computed the MP2 energies in bases up to
aug-cc-pVQZ and $\Delta$CCSD(T) in aug-cc-pVDZ,  which gives 
$-1.54$ ($-3.37$, $1.83$) kcal/mol.  
If one extrapolates the MP2 TZ-QZ results of Ref.~\onlinecite{Sinnokrot:02}
using the $X^{-3}$ extrapolation scheme,
one obtains the value of $-3.45$ kcal/mol,  not much different from the
best calculated result.
Sinnokrot {\em et al.\/} computed the MP2 energy using also
the so-called MP2-R12 explicitly correlated basis and their recommended
interaction energy is $-1.81$ ($-3.64$, $1.83$) kcal/mol.
The fairly large discrepancy between the extrapolated and MP2-R12 results
could be due to an insufficient convergence of the  resolution
of identity applied in the MP2-R12 approach in the [$spdf/spd$] bases.
Based on these
considerations, one has to assume that the uncertainty of the CCSD(T)
curve in Fig.~\ref{fig:benzene} is about $\pm 0.2$ kcal/mol.
 
Figure~\ref{fig:benzene} shows that the agreement
of SAPT(DFT) with the CCSD(T) results is very good,  in particular
taking into account the uncertainty of the latter.
In fact,  in view of this uncertainty,
the SAPT(DFT) results provide an independent set of the best
current estimates of the exact interaction energies for the benzene dimer.
The vdW-DF 
curve is deeper by about a factor of 1.4 and the minimum is
shifted by about $0.2$~\AA.
Figure~\ref{fig:benzene} includes also
the MP2 results from Ref.~\onlinecite{Tsuzuki:02},
the DFT results of Lilienfeld {\em et al.\/} from
Ref.~\onlinecite{Lilienfeld:04}, and the MP2 results of 
Sinnokrot {\em et al.\/}~\cite{Sinnokrot:02} which were used to
calibrate 
the Lilienfeld {\em et al.\/} DFT functional.
The differences between the two MP2 curves are consistent
with the sizes of the basis set effects discussed above.
For the benzene dimer,
the MP2 level of theory is not adequate and therefore the results
of Lilienfeld {\em et al.\/}~\cite{Lilienfeld:04} are very far from
the CCSD(T) benchmark.  If these authors had chosen to fit their functional
to the CCSD(T) level of theory,  the calculations would have become
significantly more expensive.
This emphasizes the fact that the DFT results of
Ref.~\onlinecite{Lilienfeld:04} can only be as accurate as the 
underlying wave-function based calculations.
Even within the MP2 model,
Fig.~\ref{fig:benzene} shows that the agreement between the DFT and MP2
energies deteriorates quickly in the regions farther from the minimum.

It should be emphasized that, in contrast to the supermolecular DFT 
approach,  SAPT(DFT) is relatively insensitive to the choice of the
DFT functional.  Several published papers show that the results given
by the former method with different choices of the functionals
can differ by an order of magnitude and can be of the wrong sign
(see, e.g., Refs.~\onlinecite{Wu:01,Tsuzuki:01}). 
The results displayed in Fig.~\ref{fig:argon} and in Table~\ref{tab:minima}
show that the SAPT(DFT) results computed with PBE0 and B97-2,  two
functionals developed using very different principles,  agree very well,
with discrepancies being of the same magnitude as other uncertainties
of the SAPT(DFT) approach.

 \begin{table}
 \caption{\label{tab:components} Interaction energy components (in kcal/mol)
 for the benzene dimer at the parallel
 ``sandwich'' geometry
 obtained with SAPT(DFT) compared to those obtained with the regular
 SAPT and a modified aug-cc-pVDZ basis in Ref.~\onlinecite{Sinnokrot:04}.}
 \begin{ruledtabular}
 \begin{tabular}{lddd}
    \multicolumn{1}{c}{} & \multicolumn{1}{c}{PBE0/3.8 \AA} & \multicolumn{1}{c}{PBE0/3.7 \AA}   & \multicolumn{1}{c}{SAPT/3.7 \AA\footnote{See Ref.~\onlinecite{Sinnokrot:04} for the SAPT components
 included at each level.}}  \\
  \hline
    electrostatic        &  0.09   &  -0.28   & -0.97                      \\
    1st order exchange   &  3.66   &  4.94    &  6.03                          \\
    induction            & -1.24   & -1.69    & -2.14
                      \\
    exchange-induction   &  1.02   & 1.46    &  1.95               \\
    dispersion           & -5.51   & -6.44   & -7.47               \\
    exchange-dispersion  &  0.50   & 0.65   &  0.94               \\
    total                & -1.48   & -1.37  & -1.66               \\
 \end{tabular}
 \end{ruledtabular}
 \end{table}

An advantage of the SAPT approach is that it gives
directly the physical components
of the interaction energy. 
In Table~\ref{tab:components}, we present such components 
for the benzene dimer and compare with SAPT calculations by
Sinnokrot and Sherrill~\cite{Sinnokrot:04}.
We have omitted in this comparison terms beyond the second order in $V$
computed in Ref~\onlinecite{Sinnokrot:04}. 
 These terms amount to $-0.14$ kcal/mol.
Generally,  the agreement between SAPT(DFT) and SAPT is good
taking into account different levels of intramonomer correlation effects,
different basis sets,  and different monomer geometries used.

In summary,  we have shown that SAPT(DFT) is capable to achieve
the accuracy of intermolecular interaction energies
similar to that of the CCSD(T)/CBS approach at a very small
fraction of computational costs.  This accuracy is much higher than that
of the vdW-DF method of Dion {\em et al.\/}~\cite{Dion:04}.
The SAPT(DFT) method
is rigorously valid for all separations between the interacting molecules.
It therefore constitutes the most accurate current approach
for practical calculations of interactions for large monomers,
containing as many as 20 atoms each.  This development will bring
some important biophysical applications within reach of computational
physics (e.g., interactions involving DNA bases,  small polypeptides,
and sugars).

\begin{acknowledgments}
The authors are grateful to Bogumi{\l} Jeziorski for reading the manuscript
and for valuable advice.
This research was supported by a grant from ARO.
\end{acknowledgments}

\bibliography{ar-kr-benzene}

\begin{thebibliography}{26}
\expandafter\ifx\csname natexlab\endcsname\relax\def\natexlab#1{#1}\fi
\expandafter\ifx\csname bibnamefont\endcsname\relax
  \def\bibnamefont#1{#1}\fi
\expandafter\ifx\csname bibfnamefont\endcsname\relax
  \def\bibfnamefont#1{#1}\fi
\expandafter\ifx\csname citenamefont\endcsname\relax
  \def\citenamefont#1{#1}\fi
\expandafter\ifx\csname url\endcsname\relax
  \def\url#1{\texttt{#1}}\fi
\expandafter\ifx\csname urlprefix\endcsname\relax\def\urlprefix{URL }\fi
\providecommand{\bibinfo}[2]{#2}
\providecommand{\eprint}[2][]{\url{#2}}

\bibitem[{\citenamefont{{X. Wu, M. C. Vargas, S. Nayak, V. L. Lotrich, and G.
  Scoles}}(2001)}]{Wu:01}
\bibinfo{author}{\bibnamefont{{X. Wu, M. C. Vargas, S. Nayak, V. L. Lotrich,
  and G. Scoles}}}, \bibinfo{journal}{J. Chem. Phys.}
  \textbf{\bibinfo{volume}{115}}, \bibinfo{pages}{8748} (\bibinfo{year}{2001}).

\bibitem[{\citenamefont{{A. D. Boese, A. Chandra, J. M. L. Martin, and D.
  Marx}}(2003)}]{Boese:03}
\bibinfo{author}{\bibnamefont{{A. D. Boese, A. Chandra, J. M. L. Martin, and D.
  Marx}}}, \bibinfo{journal}{J. Chem. Phys.} \textbf{\bibinfo{volume}{119}},
  \bibinfo{pages}{5965} (\bibinfo{year}{2003}).

\bibitem[{\citenamefont{{O. A. von Lilienfeld, I. Tavernelli, U. Rothlisberger,
  and D. Sebastiani}}(2004)}]{Lilienfeld:04}
\bibinfo{author}{\bibnamefont{{O. A. von Lilienfeld, I. Tavernelli, U.
  Rothlisberger, and D. Sebastiani}}}, \bibinfo{journal}{Phys. Rev. Lett.}
  \textbf{\bibinfo{volume}{93}}, \bibinfo{pages}{153004}
  (\bibinfo{year}{2004}).

\bibitem[{\citenamefont{Dion et~al.}(2004)\citenamefont{Dion, Rydberg,
  {Schr\"{o}der}, Langreth, and Lundqvist}}]{Dion:04}
\bibinfo{author}{\bibfnamefont{M.}~\bibnamefont{Dion}},
  \bibinfo{author}{\bibfnamefont{H.}~\bibnamefont{Rydberg}},
  \bibinfo{author}{\bibfnamefont{E.}~\bibnamefont{{Schr\"{o}der}}},
  \bibinfo{author}{\bibfnamefont{D.~C.} \bibnamefont{Langreth}},
  \bibnamefont{and} \bibinfo{author}{\bibfnamefont{B.~I.}
  \bibnamefont{Lundqvist}}, \bibinfo{journal}{Phys. Rev. Lett.}
  \textbf{\bibinfo{volume}{92}}, \bibinfo{pages}{246401}
  (\bibinfo{year}{2004}).

\bibitem[{\citenamefont{Misquitta and Szalewicz}(2002)}]{Misquitta:02}
\bibinfo{author}{\bibfnamefont{A.~J.} \bibnamefont{Misquitta}}
  \bibnamefont{and}
  \bibinfo{author}{\bibfnamefont{K.}~\bibnamefont{Szalewicz}},
  \bibinfo{journal}{Chem. Phys. Lett.} \textbf{\bibinfo{volume}{357}},
  \bibinfo{pages}{301} (\bibinfo{year}{2002}).

\bibitem[{\citenamefont{Misquitta et~al.}(2003)\citenamefont{Misquitta,
  Jeziorski, and Szalewicz}}]{Misquitta:03}
\bibinfo{author}{\bibfnamefont{A.~J.} \bibnamefont{Misquitta}},
  \bibinfo{author}{\bibfnamefont{B.}~\bibnamefont{Jeziorski}},
  \bibnamefont{and}
  \bibinfo{author}{\bibfnamefont{K.}~\bibnamefont{Szalewicz}},
  \bibinfo{journal}{Phys. Rev. Lett.} \textbf{\bibinfo{volume}{91}},
  \bibinfo{pages}{033201} (\bibinfo{year}{2003}).

\bibitem[{\citenamefont{{A. Hesselmann and G. Jansen}}()}]{Hesselmann:02-03}
\bibinfo{author}{\bibnamefont{{A. Hesselmann and G. Jansen}}},
  \bibinfo{howpublished}{Chem. Phys. Lett. {\bf 357}, 464 (2002); {\bf 362},
  319 (2002); {\bf 367}, 778 (2003).}

\bibitem[{\citenamefont{Williams and Chabalowski}(2001)}]{Williams:01}
\bibinfo{author}{\bibfnamefont{H.~L.} \bibnamefont{Williams}} \bibnamefont{and}
  \bibinfo{author}{\bibfnamefont{C.~F.} \bibnamefont{Chabalowski}},
  \bibinfo{journal}{J. Phys. Chem. A} \textbf{\bibinfo{volume}{105}},
  \bibinfo{pages}{646} (\bibinfo{year}{2001}).

\bibitem[{\citenamefont{{B. Jeziorski, R. Moszynski, and K.
  Szalewicz}}(1994)}]{Jeziorski:94}
\bibinfo{author}{\bibnamefont{{B. Jeziorski, R. Moszynski, and K. Szalewicz}}},
  \bibinfo{journal}{Chem. Rev.} \textbf{\bibinfo{volume}{94}},
  \bibinfo{pages}{1887} (\bibinfo{year}{1994}).

\bibitem[{\citenamefont{Misquitta and Szalewicz}()}]{Misquitta:04}
\bibinfo{author}{\bibfnamefont{A.~J.} \bibnamefont{Misquitta}}
  \bibnamefont{and}
  \bibinfo{author}{\bibfnamefont{K.}~\bibnamefont{Szalewicz}},
  \bibinfo{note}{{submitted to J. Chem. Phys.}}

\bibitem[{\citenamefont{Misquitta et~al.}()\citenamefont{Misquitta, Podeszwa,
  Jeziorski, and Szalewicz}}]{Misquitta:05}
\bibinfo{author}{\bibfnamefont{A.~J.} \bibnamefont{Misquitta}},
  \bibinfo{author}{\bibfnamefont{R.}~\bibnamefont{Podeszwa}},
  \bibinfo{author}{\bibfnamefont{B.}~\bibnamefont{Jeziorski}},
  \bibnamefont{and}
  \bibinfo{author}{\bibfnamefont{K.}~\bibnamefont{Szalewicz}},
  \bibinfo{note}{{manuscript in preparation}}.

\bibitem[{CAD()}]{CADPACshort}
\emph{\bibinfo{title}{{{\em CADPAC}: The Cambridge Analytic Derivatives Package
  Issue 6, {\em Cambridge, 1995. A suite of quantum chemistry programs
  developed by R. D. Amos with contributions from I. L. Alberts} et al.}}}

\bibitem[{\citenamefont{Podeszwa and Szalewicz}()}]{Podeszwa:04}
\bibinfo{author}{\bibfnamefont{R.}~\bibnamefont{Podeszwa}} \bibnamefont{and}
  \bibinfo{author}{\bibfnamefont{K.}~\bibnamefont{Szalewicz}},
  \bibinfo{note}{{work in progress}}.

\bibitem[{\citenamefont{Woon and T.H.~Dunning}(1993)}]{Woon:93}
\bibinfo{author}{\bibfnamefont{D.}~\bibnamefont{Woon}} \bibnamefont{and}
  \bibinfo{author}{\bibfnamefont{J.}~\bibnamefont{T.H.~Dunning}},
  \bibinfo{journal}{J. Chem. Phys.} \textbf{\bibinfo{volume}{98}},
  \bibinfo{pages}{1358} (\bibinfo{year}{1993}).

\bibitem[{\citenamefont{Bukowski et~al.}(1999)\citenamefont{Bukowski,
  Szalewicz, and Chabalowski}}]{Bukowski:99}
\bibinfo{author}{\bibfnamefont{R.}~\bibnamefont{Bukowski}},
  \bibinfo{author}{\bibfnamefont{K.}~\bibnamefont{Szalewicz}},
  \bibnamefont{and}
  \bibinfo{author}{\bibfnamefont{C.}~\bibnamefont{Chabalowski}},
  \bibinfo{journal}{J. Phys. Chem. A} \textbf{\bibinfo{volume}{103}},
  \bibinfo{pages}{7322} (\bibinfo{year}{1999}).

\bibitem[{\citenamefont{Williams et~al.}(1995)\citenamefont{Williams, Mas,
  Szalewicz, and Jeziorski}}]{Williams:95}
\bibinfo{author}{\bibfnamefont{H.~L.} \bibnamefont{Williams}},
  \bibinfo{author}{\bibfnamefont{E.~M.} \bibnamefont{Mas}},
  \bibinfo{author}{\bibfnamefont{K.}~\bibnamefont{Szalewicz}},
  \bibnamefont{and}
  \bibinfo{author}{\bibfnamefont{B.}~\bibnamefont{Jeziorski}},
  \bibinfo{journal}{J. Chem. Phys.} \textbf{\bibinfo{volume}{103}},
  \bibinfo{pages}{7374} (\bibinfo{year}{1995}).

\bibitem[{\citenamefont{Tsuzuki et~al.}(2002)\citenamefont{Tsuzuki, Honda,
  Mikami, and Tanabe}}]{Tsuzuki:02}
\bibinfo{author}{\bibfnamefont{S.}~\bibnamefont{Tsuzuki}},
  \bibinfo{author}{\bibfnamefont{K.}~\bibnamefont{Honda}},
  \bibinfo{author}{\bibfnamefont{M.}~\bibnamefont{Mikami}}, \bibnamefont{and}
  \bibinfo{author}{\bibfnamefont{K.}~\bibnamefont{Tanabe}},
  \bibinfo{journal}{J. Am. Chem. Soc.} \textbf{\bibinfo{volume}{124}},
  \bibinfo{pages}{104} (\bibinfo{year}{2002}).

\bibitem[{\citenamefont{Adamo et~al.}(1999)\citenamefont{Adamo, Cossi, and
  Barone}}]{Adamo:99}
\bibinfo{author}{\bibfnamefont{C.}~\bibnamefont{Adamo}},
  \bibinfo{author}{\bibfnamefont{M.}~\bibnamefont{Cossi}}, \bibnamefont{and}
  \bibinfo{author}{\bibfnamefont{V.}~\bibnamefont{Barone}},
  \bibinfo{journal}{J. Mol. Struct. (Theochem)} \textbf{\bibinfo{volume}{493}},
  \bibinfo{pages}{245} (\bibinfo{year}{1999}).

\bibitem[{\citenamefont{Wilson et~al.}(2001)\citenamefont{Wilson, Bradley, and
  Tozer}}]{Wilson:01}
\bibinfo{author}{\bibfnamefont{P.~J.} \bibnamefont{Wilson}},
  \bibinfo{author}{\bibfnamefont{T.~J.} \bibnamefont{Bradley}},
  \bibnamefont{and} \bibinfo{author}{\bibfnamefont{D.~J.} \bibnamefont{Tozer}},
  \bibinfo{journal}{J. Chem. Phys.} \textbf{\bibinfo{volume}{115}},
  \bibinfo{pages}{9233} (\bibinfo{year}{2001}).

\bibitem[{\citenamefont{Patkowski et~al.}()\citenamefont{Patkowski, Murdachaew,
  Fou, and Szalewicz}}]{Patkowski:04}
\bibinfo{author}{\bibfnamefont{K.}~\bibnamefont{Patkowski}},
  \bibinfo{author}{\bibfnamefont{G.}~\bibnamefont{Murdachaew}},
  \bibinfo{author}{\bibfnamefont{C.-M.} \bibnamefont{Fou}}, \bibnamefont{and}
  \bibinfo{author}{\bibfnamefont{K.}~\bibnamefont{Szalewicz}},
  \bibinfo{note}{{Mol. Phys., in press.}}

\bibitem[{\citenamefont{Aziz}(1993)}]{Aziz:93}
\bibinfo{author}{\bibfnamefont{R.~A.} \bibnamefont{Aziz}}, \bibinfo{journal}{J.
  Chem. Phys.} \textbf{\bibinfo{volume}{99}}, \bibinfo{pages}{4518}
  (\bibinfo{year}{1993}).

\bibitem[{\citenamefont{Dham et~al.}(1989)\citenamefont{Dham, Allnatt, Meath,
  and Aziz}}]{Dham:89}
\bibinfo{author}{\bibfnamefont{A.~K.} \bibnamefont{Dham}},
  \bibinfo{author}{\bibfnamefont{A.~R.} \bibnamefont{Allnatt}},
  \bibinfo{author}{\bibfnamefont{W.~J.} \bibnamefont{Meath}}, \bibnamefont{and}
  \bibinfo{author}{\bibfnamefont{R.~A.} \bibnamefont{Aziz}},
  \bibinfo{journal}{Mol. Phys.} \textbf{\bibinfo{volume}{67}},
  \bibinfo{pages}{1291} (\bibinfo{year}{1989}).

\bibitem[{\citenamefont{{Slav\'{\i}\v{c}ek}
  et~al.}(2003)\citenamefont{{Slav\'{\i}\v{c}ek}, Kalus, {Pa\v{s}ka},
  {Odv\'{a}rkov\'{a}}, Hobza, and {Malijevsk\'{y}}}}]{Slavicek:03}
\bibinfo{author}{\bibfnamefont{P.}~\bibnamefont{{Slav\'{\i}\v{c}ek}}},
  \bibinfo{author}{\bibfnamefont{R.}~\bibnamefont{Kalus}},
  \bibinfo{author}{\bibfnamefont{P.}~\bibnamefont{{Pa\v{s}ka}}},
  \bibinfo{author}{\bibfnamefont{I.}~\bibnamefont{{Odv\'{a}rkov\'{a}}}},
  \bibinfo{author}{\bibfnamefont{P.}~\bibnamefont{Hobza}}, \bibnamefont{and}
  \bibinfo{author}{\bibfnamefont{A.}~\bibnamefont{{Malijevsk\'{y}}}},
  \bibinfo{journal}{J. Chem. Phys.} \textbf{\bibinfo{volume}{119}},
  \bibinfo{pages}{2102} (\bibinfo{year}{2003}).

\bibitem[{\citenamefont{Sinnokrot et~al.}(2002)\citenamefont{Sinnokrot, Valeev,
  and Sherrill}}]{Sinnokrot:02}
\bibinfo{author}{\bibfnamefont{M.~O.} \bibnamefont{Sinnokrot}},
  \bibinfo{author}{\bibfnamefont{E.~F.} \bibnamefont{Valeev}},
  \bibnamefont{and} \bibinfo{author}{\bibfnamefont{C.~D.}
  \bibnamefont{Sherrill}}, \bibinfo{journal}{J. Am. Chem. Soc.}
  \textbf{\bibinfo{volume}{124}}, \bibinfo{pages}{10887}
  (\bibinfo{year}{2002}).

\bibitem[{\citenamefont{{S. Tsuzuki and H. P. L\"{u}thi}}(2001)}]{Tsuzuki:01}
\bibinfo{author}{\bibnamefont{{S. Tsuzuki and H. P. L\"{u}thi}}},
  \bibinfo{journal}{J. Chem. Phys.} \textbf{\bibinfo{volume}{114}},
  \bibinfo{pages}{3949} (\bibinfo{year}{2001}).

\bibitem[{\citenamefont{Sinnokrot and Sherrill}(2004)}]{Sinnokrot:04}
\bibinfo{author}{\bibfnamefont{M.~O.} \bibnamefont{Sinnokrot}}
  \bibnamefont{and} \bibinfo{author}{\bibfnamefont{C.~D.}
  \bibnamefont{Sherrill}}, \bibinfo{journal}{J. Am. Chem. Soc.}
  \textbf{\bibinfo{volume}{126}}, \bibinfo{pages}{7690} (\bibinfo{year}{2004}).

\end{thebibliography}

\end{document}